# Complexity Analysis of Chaos and Other Fluctuating Phenomena

Jamieson Brechtl[a], Xie Xie[b], Karen A. Dahmen[c], and Peter K. Liaw[b]

a: The Bredesen Center for Interdisiplinary Research and Education, The University of Tennessee, Knoxville TN 37996, USA

b: Department of Materials Science and Engineering, The University of Tennessee, Knoxville, TN 37996, USA

c: Department of Physics, University of Illinois at Urbana-Champaign, Champaign, IL 61820, USA

*Abstract*

The refined composite multiscale-entropy algorithm was applied to the time-dependent behavior of the Weierstrass functions, colored noise, and Logistic map to provide fresh insight into the dynamics of these fluctuating phenomena. For the Weierstrass function, the complexity of fluctuations was found to increase with respect to the fractional dimension, *D*, of the graph. Additionally, the sample-entropy curves increased in an exponential fashion with increasing *D*. This increase in the complexity was found to correspond to a rising amount of irregularities in the oscillations. In terms of the colored noise, the complexity of the fluctuations was found to be highest for the 1/f noise (f is the frequency of the generated noise) which is in agreement with findings in the literature. Moreover, the sample-entropy curves exhibited a decreasing trend for noise when the spectral exponent, *β*, was less than 1 and obeyed an increasing trend when *β* > 1. Importantly, a direct relationship was observed between the power-law exponents for the curves and the spectral exponents of the noise. For the logistic map, a correspondence was observed between the complexity maps and its bifurcation diagrams. Specifically, the map of the sample-entropy curves was negligible when the bifurcation parameter, *R*, varied between 3 – 3.5. Beyond these values, the curves attained non-zero values that increased with increasing *R*, in general.

## 1. Introduction

A variety of the sample-entropy (Sample En.) techniques have been proposed to study the complexity of time-series data representing nonlinear dynamical systems [1]. One such technique is the ApEn algorithm [2], which measures the probability that similar sequences (for a given number of points) will remain like each other when an additional point is added. However, this method contains bias due to self-matching. To overcome this issue, the SampEn technique, which excludes self-matching in the calculation, was proposed by Richman et al. [3]. Here the SampEn is defined as the negative natural logarithm of the conditional probability that two sequences remain similar at the next point.

The multiscale entropy (MSE) algorithm was proposed by Costa et al. [4] to calculate SampEn over a range of scales to represent the complexity of a time series. Importantly, the MSE algorithm resolved an issue with the ApEn method, which stated that the white noise consisted of fluctuations that were more complex than those associated with the 1/f noise [5]. Here, f is defined as the frequency of the generated noise, which is bounded between arbitrarily small and large values. However, this result was contradictory since the 1/f noise was thought to be more intricate in nature. However, the MSE technique, as proposed by Costa et al., showed that although the white noise was more complex at lower scales, the 1/f noise possessed higher levels of complexity at larger scaling factors [4, 6].

In addition, the MSE algorithm has been found to be useful in analyzing and modeling temporal data, such as serrated flow, during mechanical deformation, in different alloy systems [7-9], physiological-time series [4, 6, 10], bearing vibration data [11], and financial time series [12, 13]. However, the MSE technique does have issues, such as problems in accuracy and validity at large scale factors [5]. To tackle these issues, Wu et al. [14] developed the composite multiscale entropy (CMSE) algorithm, which can estimate the complexity more accurately but increases the chance of producing undefined values. This technique has since been used to analyze financial-time series [15, 16].

More recently, Wu et al. modified the CMSE algorithm slightly to produce what is known as the refined composite multiscale entropy (RCMSE) algorithm [5]. In their work, they compared the complexity of the white and 1/f noise. In terms of accuracy, it was found that the RCMSE algorithm outperformed both the MSE and CMSE algorithms. Like its predecessors, this technique has been used to study the complexity of physiological systems [17], the serrated flow in different alloy systems [18, 19], and the intrinsic dynamics of traffic signals [20].

Therefore, the goal of the present work is to use the RCMSE method to model and analyze the complexity of different fluctuating phenomena. These phenomena include the colored noise, the Weierstrass function, and the logistic map. In terms of the colored noise, the current study will expand upon the studies conducted by [4, 6, 9, 21] on white and 1/f noise, where the noise with spectral exponents ranging from – 2 to 2 will be modeled and analyzed. Furthermore, this study will provide an innovative way to understand how the regularity of a fractal function changes with respect to its fractional dimension. This investigation also takes an original approach to examining the logistic map, where the complexity of its fluctuations will be examined with respect to its chaotic behavior. Therefore, this work is significant since it advances our understanding of the above phenomenon.

## *2. Refined Multiscale Entropy Modeling and Analysis*

For this section, the methodology of [5] will be used. Given a discrete time series of the form, $X = [\ x_1\ x_2\ ...\ x_i\ ...\ x_N\ ]$, one constructs the coarse-grained (averaged) time series, $y_{j,k}^\tau$, using Equation (1), which is written as:

$$y_{k,j}^{\tau} = \frac{1}{\tau} \sum_{i=(j-1)\tau+k}^{j\tau+k-1} x_i \quad ; \quad 1 \leq j \leq \frac{N}{\tau} \quad 1 \leq k \leq \tau \tag{1}$$

Here $N$ is the total number of points in the original data set, and $k$ is an indexing factor, which dictates at which $x_i$ one begins the coarse-graining procedure. Additionally, one should note that the coarse-grained series, $y_{1,1}^1$, is simply the original time series, $X$. Figure 1 gives a schematic illustration of the coarse-graining procedure. At this point, one constructs the template vectors, $\boldsymbol{y}_i^{\tau,m}$, of dimension, $m$ [4]:

$$\boldsymbol{y}_i^{\tau,m} = \{y_i^\tau \; y_{i+1}^\tau \; .... \; y_j^\tau \; .... \; y_{i+m-1}^\tau\} \quad ; \quad 1 \leq i \leq N-m \tag{2}$$

Once $y_{k,j}^\tau$ is constructed, the next step is to write the time series of $y_k^\tau$ as a vector for each scale factor, $\tau$:

$$\boldsymbol{y}_k^\tau = \{y_{k,1}^\tau \; y_{k,2}^\tau \; .... \; y_{k,N}^\tau\} \tag{3}$$

The next step in the process is to find $n$ matching sets of distinct template vectors. It should be noted that the previous studies used $m = 2$ as the size of the template vector [3-5]. For two vectors to match, the infinity norm, $d_{jk}^{\tau,m}$, of the difference between them must be less than a predefined tolerance value, $r$. Here the infinity norm may be written as:

$$d_{jk}^{\tau,m} = \|\boldsymbol{y}_j^{\tau,m} - \boldsymbol{y}_k^{\tau,m}\|_\infty = \max\{|y_{1,j}^\tau - y_{1,k}^\tau| \; ... \; |y_{i+m-1,j}^\tau - y_{i+m-1,k}^\tau|\} < r \tag{4}$$

Typically, r is chosen as $0.1 - 0.2$ times the standard deviation, of the original data set [6]. This choice ensures that the sample entropy relies on the sequential ordering, and not the variance, of the original time series. For this study, a value of $r = 0.15\sigma$ will be used.

Figure 2 illustrates the matching process for the coarse-grained series, $y_{1,j}^1 = X(j)$ (here $k = 1$) [6]. In the graph, there is the template sequence, $\{x(1), x(2), x(3)\}$, which matches the template sequence, $\{x(28), x(29), x(30)\}$, meaning that there is a matching three-component template set. Here the matching points for the three-component templates are denoted by blue boxes in the figure. This calculation is, then, repeated for the next three-component template sequence in which a total count of matching template sequences is taken. Then the entire process is repeated for all two-component template sequences. The number of matching two- and three component template sequences are again summed and added to the cumulative total.

This procedure is performed for each $k$ from 1 to $\tau$ and, then, the number of matching template sequences, $n_k^m$ and $n_k^{m+1}$, is summed, which is written as:

$$RCMSE(y, \tau, m, r) = Ln\left(\frac{\sum_{k=1}^{\tau} n_{k,\tau}^m}{\sum_{k=1}^{\tau} n_{k,\tau}^{m+1}}\right) \quad (5)$$

The RCMSE value is typically denoted as the sample entropy of sample en. for short. As with other techniques, the RCMSE curves are used to compare the relative complexity of normalized time series [6]. However, an advantage of the RCMSE method is that it has a lower chance of inducing undefined entropy as compared to earlier algorithms [5]. As was done in previous studies [4-6], the sample entropy was plotted for scale factor values ranging from 1 to 20.

## 3. Modeling and Analysis

### 3.1 Weierstrass Functions

Weierstrass functions are an example of a function which is continuous but differentiable nowhere [22]. A proof of the non-differentiability of this function can be found in [23], and a discussion as to its fractal nature can be read in [24]. Typically, the Weierstrass function has a similar form to the following [25]:

$$W(t) = \sum_{k=1}^{\infty} \frac{e^{i(\gamma^k t + \varphi_k)}}{\gamma^{(2-D)k}} \qquad (6)$$

where $D$ is the fractional dimension with $1 < D < 2$, $\gamma > 1$, and $\varphi_k$ is an arbitrary phase. Here, the real and imaginary parts of Equation (6) are known as the Weierstrass cosine and sine functions, respectively. Additionally, $D$ will be termed as the fractional dimension to avoid technical arguments over which type of dimension, $D$, represents, such as the box-counting dimension, fractal dimension, or the Hausdorff-Besicovitch dimension [25].

Although Weierstrass functions cannot be differentiated in the conventional sense, they have been shown to be differentiable to fractional order [26-30]. Furthermore, both integrating and differentiating functions to arbitrary order involve more generalized definitions, as compared to those found in the integer order calculus. For example, the fractional integral has been defined as [31]:

$$_cD_t^{-\alpha} f(t) = \frac{1}{\Gamma(\alpha)} \int_c^t (t-t')^{\alpha-1} f(t') dt' \quad Re\ \alpha > 0 \qquad (7)$$

Here $\Gamma$ is the well-known gamma function, and $\alpha$ is the order of the derivative, which extends across the positive reals. Expanding upon Equation (7), Oldham and Spanier show that the fractional derivative of a function, $f(t)$, may be written as [32]:

$$_aD_t^{\alpha} f(t) = \frac{d^n}{dt^n} {_aD_t^{\alpha-n}} f(t) = \frac{1}{\Gamma(n-\alpha)} \frac{d^n}{dt^n} \int_a^t (t-t')^{n-\alpha-1} f(t') dt' \quad Re\ \alpha > 0 \qquad (8)$$

In the spirit of the work found in [26], we take the fractional integral, as defined in Equation (7) and apply it to the righthand side (r. h. s.) of Equation (6), while taking the limit of $c \to -\infty$ [from Equation (7)]:

$$_{-\infty}D_t^{-\alpha}W(t) = \frac{1}{\Gamma(\alpha)}\sum_{k=1}^{\infty}\frac{e^{i\varphi_k}}{\gamma^{(2-D)k}}\int_{-\infty}^{t}\frac{e^{i\gamma^k t'}}{(t-t')^{1-\alpha}}\,dt' \quad 0 < \alpha < 1 \tag{9}$$

Applying the substitution twice and integrating yields:

$$_{-\infty}D_t^{-\alpha}W(t) = \sum_{k=1}^{\infty}\frac{e^{i\left(\gamma^k t + \varphi_k - \frac{\pi\alpha}{2}\right)}}{\gamma^{[2-(D-\alpha)]k}} \tag{10}$$

In a similar fashion, we solve for the fractional derivative of the $W(t)$:

$$_{-\infty}D_t^{\alpha}W(t) = \frac{1}{\Gamma(1-\alpha)}\sum_{k=1}^{\infty}\frac{e^{i\varphi_k}}{\gamma^{(2-D)k}}\frac{d}{dt}\int_{-\infty}^{t}\frac{e^{i\gamma^k t'}}{(t-t')^{\alpha}}\,dt'$$

$$= \sum_{k=1}^{\infty}\frac{e^{i\left(\gamma^k t + \varphi_k + \frac{\pi\alpha}{2}\right)}}{\gamma^{[2-(D+\alpha)]k}} \quad ; \quad 0 < \alpha < 1 \tag{11}$$

For the present work, only the fractional integral and derivatives for the cosine series (real part) of Equation (6), denoted as $W_c(t)$, will be analyzed. Additionally, $W_c(t)$ was determined by summing the first 20 terms of the series. The fractional integral and derivative for $W_c(t)$ can be written as:

$$_{-\infty}D_t^{-\alpha}W_c(t) = \sum_{k=1}^{\infty}\frac{\cos\left(\gamma^k t + \varphi_k - \frac{\pi\alpha}{2}\right)}{\gamma^{[2-(D-\alpha)]k}} \tag{12}$$

$$_{-\infty}D_t^{\alpha}W_c(t) = \sum_{k=1}^{\infty}\frac{\cos\left(\gamma^k t + \varphi_k + \frac{\pi\alpha}{2}\right)}{\gamma^{[2-(D+\alpha)]k}} \tag{13}$$

Before moving on, a few things should be discussed. It can be seen from Equations (10)-(13) that the fractional derivative and integral of *W(t)* is simply another Weierstrass function, which possesses both a different fractional dimension D' (D' = D ± α) and a complex exponential, which has undergone a phase shift. As noted in [26], the fractional integral of the Weierstrass function decreases the fractional dimension by a factor equal to the order of the derivative. In contrast, the fractional derivative increases the fractional dimension of the function by the same amount.

An alternate way to fractionally-differentiate the Weierstrass function was discussed in [29, 30]. In their work, they differentiated the cosine and sine series term-by-term, using the power-law rule for fractional derivatives [32, 33]:

$$_0D_t^\alpha W_c(t) = {_0D_t^\alpha} \sum_{k=1}^{\infty} \frac{\cos(\gamma^k t + \varphi_k)}{\gamma^{(2-D)k}} = \sum_{k=1}^{\infty} \frac{C_t(-\alpha, \gamma^k)}{\gamma^{(2-D)k}} = \sum_{k=1}^{\infty} \frac{t^{-\alpha} E_{2,1-\alpha}(-\gamma^{2k} t^2)}{\gamma^{(2-D)k}} \quad (14)$$

Here $C_t(-\alpha,\gamma)$ is as defined in [31], while $E_{\delta,\theta}(-t)$ is simply the two-parameter Mittag-Leffler function [34]. Both positive and negative values for *α* can be applied to the above equation. In addition, the lower limit for the derivative was set to 0. For the purposes of simplification, it was assumed that $\varphi_k = 0$.

## *3.2 Colored Noise*

The colored noise analyzed in the present work was made, using a similar method, as that in [4]. Here 200 sets of the uniformly-distributed white noise signal composed of $10^4$ points were generated. Each set was then fast Fourier transformed in which the resulting power spectrum was filtered to behave according to a $1/f^\beta$ distribution. To obtain the desired waveform, the resulting data was inverse Fourier transformed. This process was done for $\beta$ values ranging from - 2 to 2 in 0.25 increments. To highlight the decreasing/increasing trend of the complexity values with respect to the scale factor, the sample entropy was plotted for scale factors ranging from 1 – 30.

## *3.3 Chaotic Systems (Logistic Map)*

The logistic map is one of the simplest examples of chaos. Since this phenomenon has been written in detail elsewhere [35-40], it will not be discussed here. In terms of its characteristics, the logistic map consists of an iterative polynomial form that is defined as:

$$x_{n+1} = Rx_n(1 - x_n) \qquad (15)$$

where $R$ is the bifurcation parameter. Typically, the range of values for the map are $0 \leq x_n \leq 1$ and $1 \leq R \leq 4$.

## *4. Results*
## *4.1 Weierstrass Functions*

Figures 3 (a)-(c) show a plot of this series (Equation (6)) for $\varphi_k = 0$ (a fractional dimension of 1.5), $\gamma = 2$ with its fractionally-integrated ($\alpha = -0.2$), and differentiated ($\alpha = 0.2$) counterparts. As can be seen in Figure 3 (a), where the curve was integrated, the curve appears less rough, as compared to the original function. In contrast, the curve in Figure 3 (c), where the function was differentiated, exhibits an increased roughness. Therefore, this change in the fractional dimension of the Weierstrass function can be intuitively understood in terms of how the shape of the graph changes. In addition, the magnitude of the function was found to increase with respect to the fractional dimension of the function. This increase in the magnitude of the function was also observed by Liang et al. [29].

It was previously claimed in [27] that the differintegral, as used in Equations (9) and (11), may not be applied to a purely-imaginary ordinary exponential, as was done in [26]. However, the current work confirmed the derivations of West et al. [25] via the substitution. Namely, the integral in the above equations were converted into gamma functions for which the final solutions were derived.

Figure 4 shows the RCMSE results for $W_c(t)$, as discussed above, with $\varphi_k = 0$ (a fractional dimension of 1.5), $\gamma = 2$, and its corresponding fractionally-integrated and differentiated ($-0.4 \leq \alpha \leq 0.4$) functions. Here each data set consists of $3 \times 10^5$ points. As can be seen, the sample entropy increases with respect to the fractional dimension of the function. In terms of the scale factor, the complexity exhibits an increasing trend for fractional dimensions ranging from 1.1 to 1.7. At $D = 1.8$, there is an initial decrease in the entropy with respect to the scale factor, followed by an increase. However, at $D = 1.9$, the sample entropy shows a decreasing trend for $\tau$. The increasing trend with respect to $D$ suggests that Weierstrass functions with this range of dimensions contain the increasing complexity at all scales. Moreover, fractionally-differentiating the Weierstrass function leads to a greater irregularity of the fluctuations at a noticeably-higher rate, as compared to when the function is integrated.

To obtain a qualitative picture as to how the variability of the Weierstrass function changes with respect to $D$, a Poincaré plot of the r. h. s. of Equations (12) and (13) was made and is shown in Figure 5. Here, the plot was made for fractional orders of $\alpha = -0.4, -0.15, 0.15$, and $0.4$, which correspond to fractional dimensions of $D = 1.1, 1.35, 1.65$, and $1.9$, respectively. As can be observed in the graph, the points begin to noticeably spread at $D = 1.65$. In addition, the separation between points becomes significantly greater for $D = 1.9$, which reveals the increasing irregularity in the data.

Figures 6 (a)-(c) show the cosine series from Equation (14) with its fractionally-integrated and differentiated counterparts (plotted for $t = 4$ to 6 s). Notice the similarities between these figures and Figures 3 (a) - (c). Figure 7 illustrates the sample entropy for Equation (14) with $\alpha$ values ranging from - 0.4 - 0.4. From the figures, one can notice almost an exact resemblance between Figures 4 and 7.

## *4.2 Colored Noise*

Figure 8 shows the mean sample entropy of the colored noise data for $-2 \leq \beta \leq 2$ and $1 \leq \tau \leq 30$ plotted with the colored noise plots for brown ($\beta = 2$), pink (or 1/f with $\beta = 1$), white ($\beta = 0$), blue ($\beta = -1$), and violet noises ($\beta = -2$). It should be noted that the results for white and 1/f noise are similar to those reported in [4, 6, 9, 21]. In addition, the sample entropy exhibits monotonic behavior (increasing and decreasing) with respect to the scale factor across all $\beta$ exponents. Furthermore, the MSE curves appear to shift from the strictly increasing to decreasing trends at the graph for the 1/f noise ($\beta = 1$). Additionally, the curves, which correspond to $\beta < -1$, have decreasing trends that end with the sample-entropy values near zero at higher scale factors. This result indicates that as a larger number of points are averaged together, the graphs become more regular. In contrast, the curves which are increasing become more complex as the average contains a greater number of points.

The sample-entropy curves from the above figure were then fit according to the power-law form of $SE(n, \beta) = C\tau^{n(\beta)}$, where $n$ is the power-law-fitted exponent, $\beta$ is the spectrum exponent, and $C$ is some coefficient. It was found that there was a linear relationship for which $n = 0.35 - 0.34\beta$ with a corresponding r-squared value of 0.99. The results indicate that there is a strong positive correlation between these two parameters. Based on the linear fit, the concavity of the sample-entropy function undergoes a sign change when the power-spectrum exponent, $\beta$, is approximately 0.97, which is relatively close to the exponent for the 1/f noise.

## *4.3 Logistic Map*

A plot for the logistic map with $R$ = 3.2, 3.6, and 3.9 in the initial condition of $x_1 = 0.1$ is shown in Figure 9. For $R$ = 3.2, $x_n$ oscillates between two values, which is characteristic of the very simple behavior. However, as can be observed in the graph, the fluctuations become more irregular as $R$ increases.

To gain a better understanding of how the irregularity of the fluctuations changes with respect to $R$, the RCMSE method was applied to Equation (15) for $3 \leq R \leq 4$ and the initial condition of $x_1 = 0.1$. Here, $R$ was given these values since this is where the interesting behavior occurs [41]. Figure 10 shows the resulting complexity values (for $1 \leq \tau \leq 20$) for this range of values. In the first region, where $3 \leq R < 3.5$, the complexity is negligible for all scale factors. For $R$ = 3.5, there are non-zero values in the curve for scale factors of 13 and 15, which are, respectively, equal to 0.22 and 0.18. For $R > 3.5$, the sample entropy begins to noticeably increase with respect to the bifurcation parameter.

Here the curves, in general, increase with respect to the bifurcation parameter. In addition, the curves initially rise with the scale factor until they reach a maximum and then begin to decrease thereafter. Moreover, the curve for $R$ = 3.85 has similar characteristics, as compared to the curves for $R$ = 3.5 and 3.55, where there are peaks at midrange scale factors. However, the peaks for the curve at $R$ = 3.85 are relatively larger in magnitude.

Figure 11 shows the bifurcation diagram for Equation (15) with $2.8 \leq R \leq 4.0$. Here the bifurcation diagram was divided into subsections (A black box denotes each subsection in Figure 11 a) in which some of the boundaries consist of bifurcation points). In addition, arrows link the subsections of the bifurcation diagram to their respective sample-entropy curves. Interestingly, as can be observed in the Figure, it appears that complexity of the fluctuations at a given R corresponds to the number of asymptotic values visited there. In this diagram, Figure 10 was rotated clockwise to obtain a better picture of how the overall behavior of the sample-entropy curves varies with respect to the bifurcation parameter.

As can be seen in the figure, the oscillations, which occur between the first and second bifurcation points, correspond to curves, which have negligible values. The above implies that the fluctuations occurring in this region are characteristic of simplistic behavior. However, between the second and third bifurcation points, which occurs for $3.45 < R \leq 3.55$, $x$ oscillates between four points. In this region, the sample-entropy curves contain some non-zero values, meaning that there is some irregularity for these types of oscillations.

Above all, once $R$ becomes greater than 3.55, the sample-entropy curves begin to rise, especially for points at lower scale factors. For the most part, this section corresponds to the region of the bifurcation diagram containing a high density of points. Nevertheless, there are some areas in the diagram, which contain no points, and appear to parallel the sample-entropy curves, which are lower in the overall magnitude, such as the curve for $R = 3.85$. Beyond the above bifurcation parameter, the curves continue to rise again until $R = 4$. Since the behavior of $x_n$ diverges for $R > 4$, the complexity was not calculated for this range.

## 5. Discussion

### 5.1 Weierstrass Functions

It was found that fractionally-differentiating the Weierstrass function resulted in an enhanced roughness of the curves while integration had the opposite effect. Furthermore, this roughness increased with respect to the fractional dimension of the curve. From the results of the complexity analysis and modeling, it was found that the degree of roughness corresponded to the irregularity of the oscillations. As observed in Figures 4 and 7, the sample entropy appears to increase in an exponential fashion with respect to the fractional dimension (and derivative order) of the curve. Moreover, the sample entropy did not exhibit the decreasing behavior with respect to $\tau$ for a given $D$, which indicates that there is complex behavior at all scales. This link between the fractional dimension and the irregularity of the oscillations of the function was also found in the work conducted by West et al. [25, 26].

Also, the increase in the complexity of the Weierstrass function, as seen in Figure 4, arises from an increase in the information content of its oscillations as $D$ rises [42]. This increase is a consequence of the growth in the unpredictability, or irregularity of the state due to more erratic fluctuations. Furthermore, the decrease in the predictability of the fluctuations for $W_c(t)$ was further observed in the Poincaré plot from Figure 5, where the separation of consecutive points accelerated for larger values of the fractional dimension.

As can be observed in Figures 3 (a)-(c) and 6 (a)-(c), the curves produced from Equations (12) and (13) were very similar, if not identical, to those fashioned from Equation (14). In addition, their sample-entropy curves, as seen in Figures 4 and 7, behaved in a similar fashion where both curves increased in a parabolic trend with respect to the order of the derivative and $D$. The similarities between the two forms were unexpected, since the former involves altering *D*, while the latter does not change this quantity. At the time of this writing, the implications of the above result are not entirely understood, but will hopefully be the subject of further investigation.

## *5.2 Colored Noise*

The results for the colored-noise modeling and analysis indicate that the sample-entropy curves, as shown in Figure 8, appear to comprise a spectrum. As can be seen, the complexity curves for noise corresponding to $1 < \beta < 2$ have increasing trends while the curves for the noise with $\beta < 1$ decrease with respect to $\tau$. In addition, the curves shift from the concave up to concave down at $\beta \approx 1$, for reasons that are not entirely understood. Furthermore, it is well known that the spectral exponent values of $0 < \beta < 2$ correspond to fluctuations that consist of the antipersistent behavior, which means that the magnitude of fluctuations tend back towards the mean of the data set [43]. Therefore, the antipersistent behavior can also be thought of as self-regulating. Moreover, the white noise, which has a power spectrum that is constant with respect to frequency ($\beta = 0$), consists of fluctuations, which are uncorrelated.

Like other work, including [4, 6, 9, 21], the results of the present modeling and analysis found that the 1/f noise, in general, consists of fluctuations that are more complex in nature, as compared to other colored noise, which can be seen Figure 8. More specifically, the sample-entropy values for the 1/f noise remain higher for $\tau > 10$, as compared to other noise. This result indicates that this type of noise is resistant to a loss in correlations at all scales. Furthermore, this resistance to a loss in complexity may explain why intricate phenomenon which is conducive to life, such as heart-beat contractions or neuronal firing, observes fluctuations, which imitate the 1/f noise.

A link between the complexity of the time-series data and heart function was observed in studies conducted by Costa et al. [4, 6]. For both studies, they found that subjects with the healthy heart function had increasing trends (concave down) in the sample entropy with respect to the scale factor. Furthermore, they reported the same increasing trend when comparing young and elderly patients that were both healthy, although the complexity was significantly higher for all scale factors in the former. In contrast, subjects that experienced atrial fibrillation exhibited decreasing trends in the complexity of the heart-beat-interval-time-series. Additionally, subjects who had the congestive heart failure had the lowest sample entropy values for $\tau > 2$, as compared to the other two conditions. Moreover, the sample entropy was higher at larger scale factors for the healthy subjects, as compared to the values for the atrial fibrillation.

The above results were subsequently compared with the data attained from this study on the modeling and analysis of the colored noise. Based on the results of [4], the entropy measure for the healthy subjects most resembled the curve from the colored noise data from Figure 8 for the spectral exponents of $1.25 < \beta < 1.50$. The model predictions are somewhat similar to the results of another study which found that $\beta$ was 1.1 for healthy individuals [44]. Importantly, the above trend indicates that the heart-beat rhythms associated with healthy patients contain fluctuations that are self-regulating to keep the heart rate away from extreme values, which is consistent with homeostasis.

On the other hand, when comparing the results of this analysis with the data for subjects with atrial fibrillation from [4, 6], the complexity values were comparable to the colored noise with $\beta$ values ranging from 0 - 0.25 (see Figure 8). This result suggests that the heart-beat patterns for people in this condition possess lower correlations, and are almost completely random in nature. In contrast, the subjects that underwent congestive heart failure had heart-beat fluctuations with sample-entropy values that varied between the curves for $1.75 < \beta < 2$. The above statement indicates that patients, which had this condition, exhibited heart-beat fluctuations that are very similar to the brown noise, as shown in Figure 8. Since the brown noise corresponds to Brownian motion, it may be thought that those who suffer from this complication exhibit heart-beat patterns that correspond to random-walk processes, which consist of trivial long-range correlations [45].

## *5.3 Logistic Map (Chaos)*

As observed in Figure 10 and 11, the complexity was found to be negligible when the bifurcation parameter ranged from 3 - 3.45. The above region corresponds to oscillations before the first bifurcation point, where the system behaves as a limit cycle attractor. More specifically, $x_i$ oscillates between two values, which is indicative of the simplistic behavior. However, the sample-entropy curves begin to contain nonzero values for $3.45 \leq R < 3.55$, which correspond to the region between the first and second bifurcation points in Figure 11. In this range, the system oscillates between 4 points, which is indicative of the behavior that is more irregular than the fluctuations of the system in the previous region.

After this second bifurcation, the complexity of the oscillations continues to rise, until the bifurcation parameter reaches 3.56995 [35], which is where the system transitions from the predictable to chaotic behavior. Beyond this point, the sample-entropy curves, in general, increase at a much faster rate with respect to the bifurcation parameter. This increase in the complex behavior of Equation 13, as $R$ ranges from 3.56995 to ~ 3.82843, is characterized by a periodic phase interrupted by bursts of the aperiodic behavior [46]. Moreover, there is a sudden decrease in the complexity at $R = 3.85$, which is most likely associated with the island of stability at $R =$ 3.82843 [47], which shows the non-chaotic behavior. Furthermore, this decrease in the complexity arises from the system oscillating between only three points.

With respect to the scale factor, the curves initially rise until they reach a maximum for $\tau \sim 5$ and, then, begin to decrease, as seen in Figure 10. This behavior simply means that the irregularity of the oscillations achieves a maximum when the data set is averaged for every 5 points. However, the underlying physics of this behavior is not well understood and will hopefully be the subject of future endeavors.

## *6. Conclusions*

To summarize, the refined composite multiscale entropy algorithm was applied to the time-dependent oscillatory behavior of Weierstrass functions, the colored noise, and the logistic map. Here, several interesting results were found. Firstly, the complexity of fluctuations for the Weierstrass cosine function were found to increase with respect to the fractional dimension of the graph. Furthermore, the sample-entropy curves increased in an exponential fashion with respect to the fractional dimension, $D$, of the graph. This increase in the complexity was found to correspond to the irregularity of the oscillations. Secondly, in terms of the colored noise, the complexity was found to be highest for the 1/f noise, which is in agreement with findings in the literature. Moreover, the sample-entropy curves exhibited a decreasing trend for noise when the spectral exponent, $\beta$, was less than 1 and obeyed an increasing trend when $\beta > 1$. Additionally, the power-law exponent for the curves had a direct correlation with the spectral exponents of the noise. For the logistic map, the sample-entropy curves were negligible when the bifurcation parameter, $R$, varied between 3 – 3.5. Beyond these values, the curves attained non-zero values that generally increased with respect to $R$, in general.

**(a)**    k = 1, τ = 2

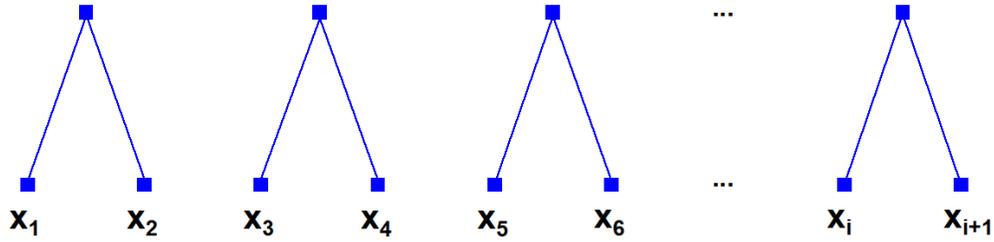

**(b)**    k = 2, τ = 2

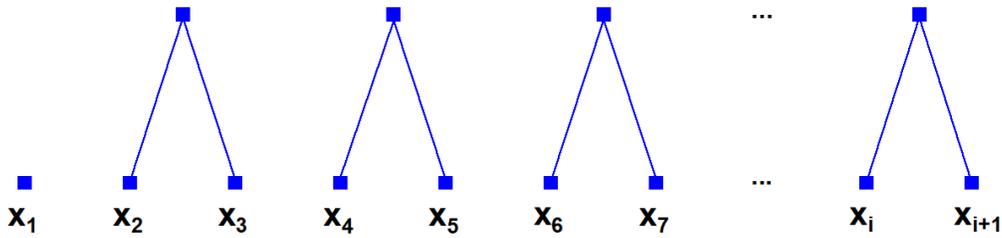

**(c)**    k = 1, τ = 3

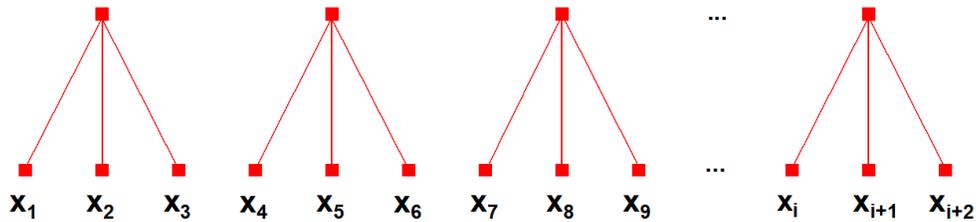

**(d)** $k = 2, \tau = 3$

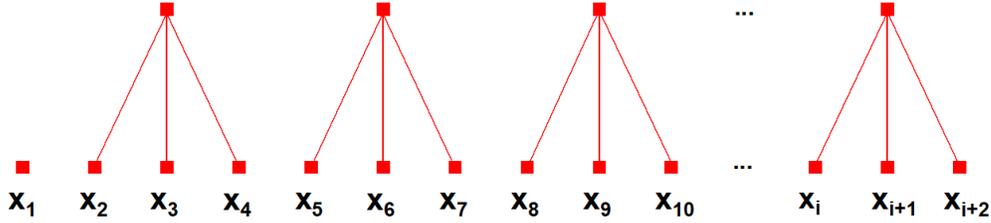

$$y^3_{2,1} = \frac{X_2 + X_3 + X_4}{3} \quad y^3_{2,2} = \frac{X_5 + X_6 + X_7}{3} \quad y^3_{2,3} = \frac{X_8 + X_9 + X_{10}}{3} \quad y^3_{2,j} = \frac{X_i + X_{i+1} + X_{i+2}}{3}$$

Figure 1: Schematic for the coarse-graining procedure for (a) $k = 1, \tau = 2$, (b) $k = 1, \tau = 3$, (c) $k = 2, \tau = 2$, and (d) $k = 2, \tau = 3$.

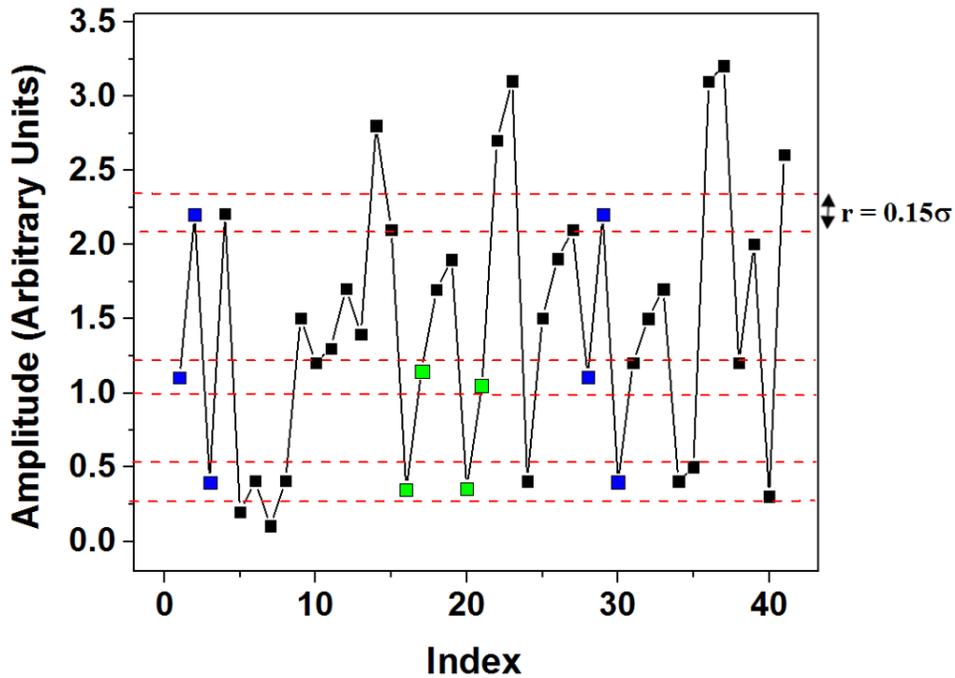

Figure 2: A simulated time series, $y^1_{1,j} = X(j)$, is shown to illustrate the procedure for calculating the sample entropy for the case, $m = 2$, and a given $r$ (which typically varies between $0.1 - 0.2\sigma$, where $\sigma$ is the standard deviation of the time series [6]). Dotted horizontal lines around data points represent $X(j) \pm r$. When the absolute difference between two data points is less than or equal to $r$, they are considered as a matching pair. A matching pair of sizes, 3 and 2, are, respectively, indicated by the blue (points $\{x(1), x(2), x(3)\}$ and $\{x(28), x(29), x(30)\}$) and green boxes in the figure.

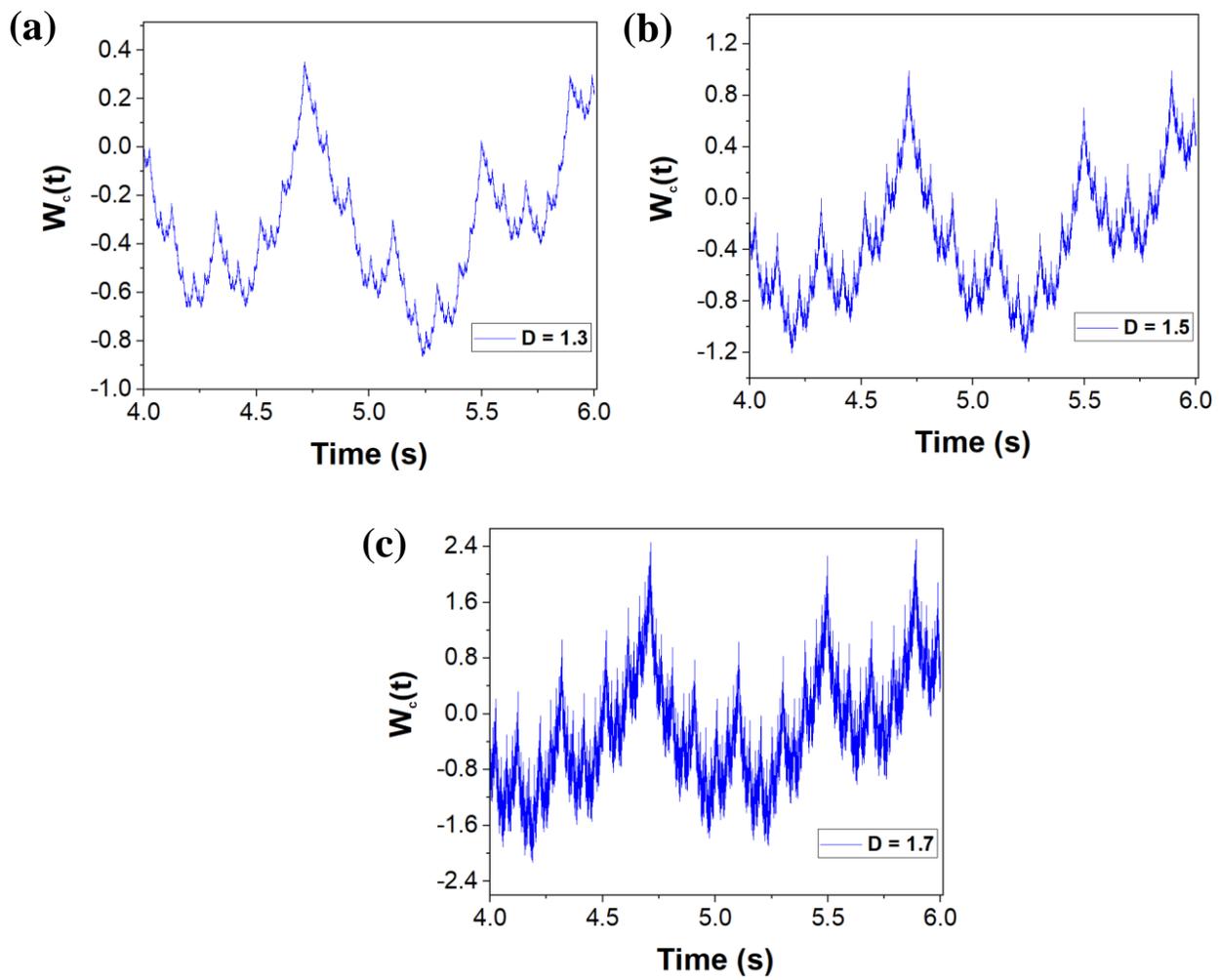

Figure 3: Plot of (a) Equation (12) with the fractional order of $\alpha = -0.2$ ($D = 1.3$), (b) the cosine series from Equation (6), and (c) Equation (13) with the fractional order of with $\alpha = 0.2$ ($D = 1.7$).

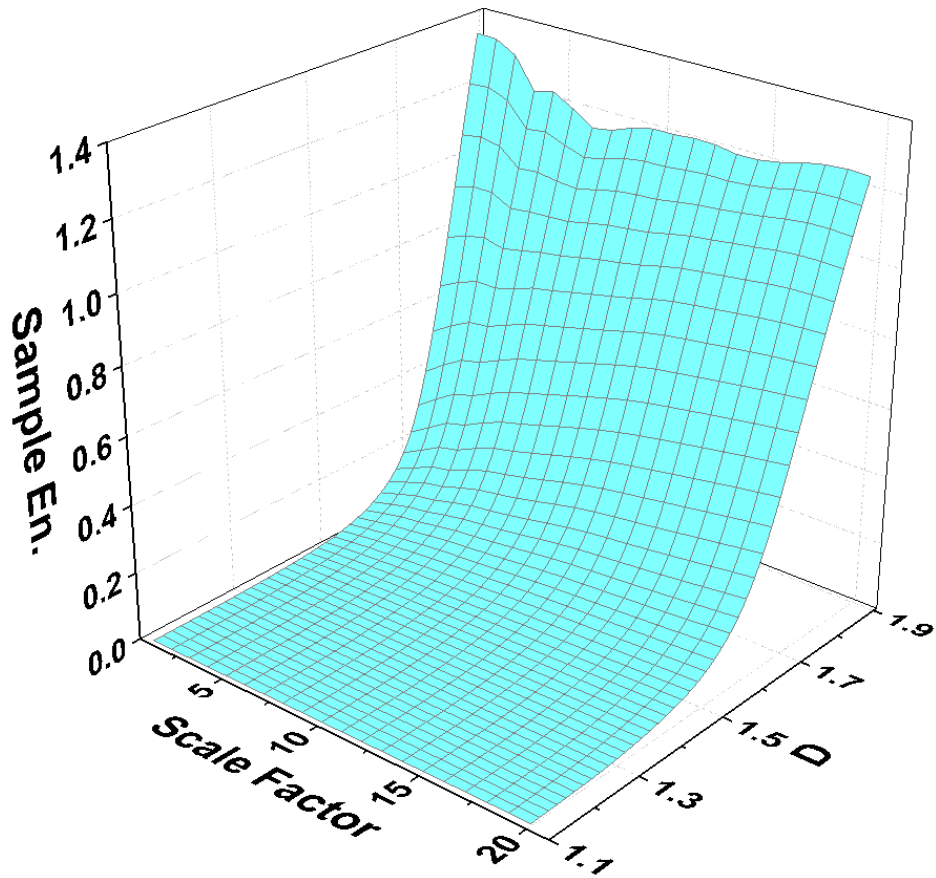

Figure 4: Sample entropy for the cosine series from Equation (6) for the fractional dimension, *D*, ranging from 1.1 – 1.9 and for - 0.4 ≤ $α$ ≤ 0.4.

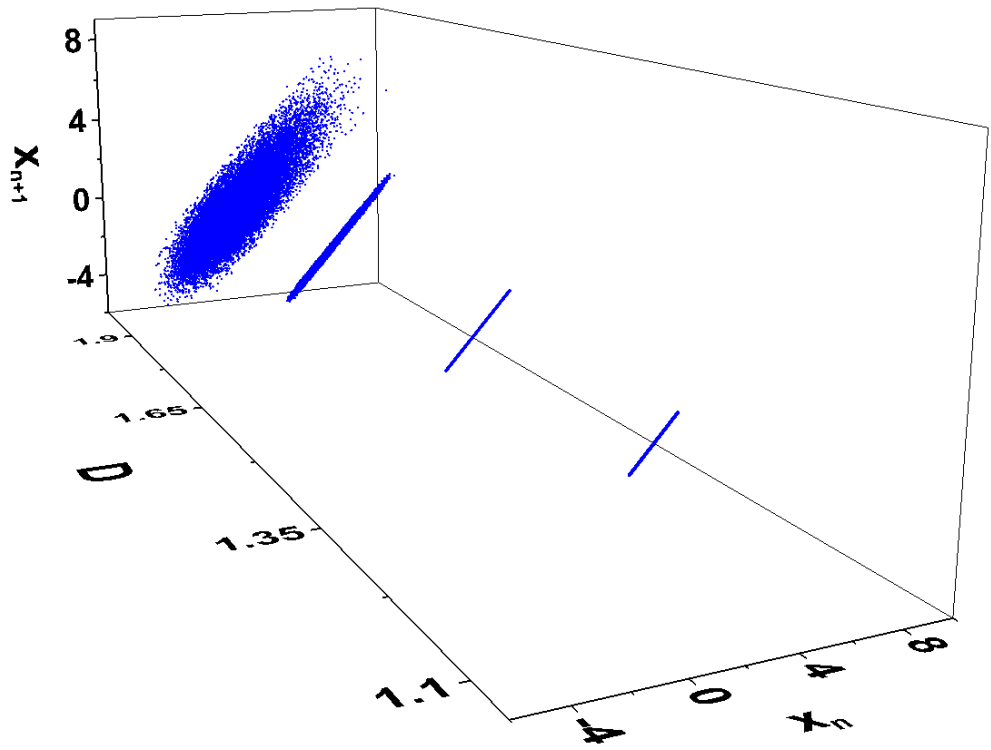

Figure 5: Poincaré plot of the r.h.s. of Equations (12) and (13) for fractional orders of $\alpha = -0.4, -0.15, 0.15$, and $0.4$, which correspond to fractional dimensions of $D = 1.1, 1.35, 1.65$, and $1.9$, respectively.

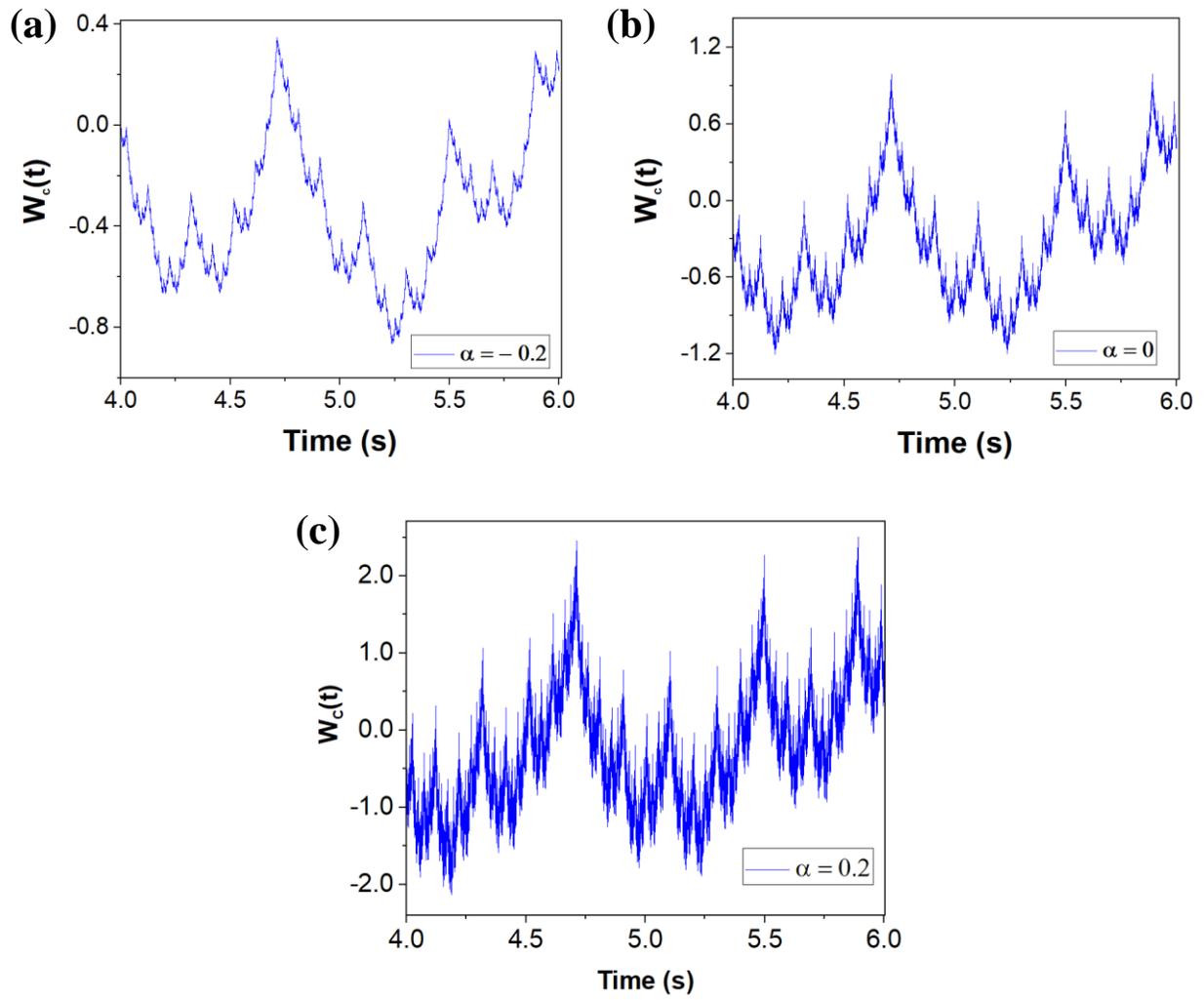

Figure 6: Plot of Equation (14) with the fractional order of (a) $\alpha = -0.2$ ($D = 1.3$), (b) $\alpha = 0$ ($D = 1.5$), and (c) $\alpha = 0.2$ ($D = 1.7$).

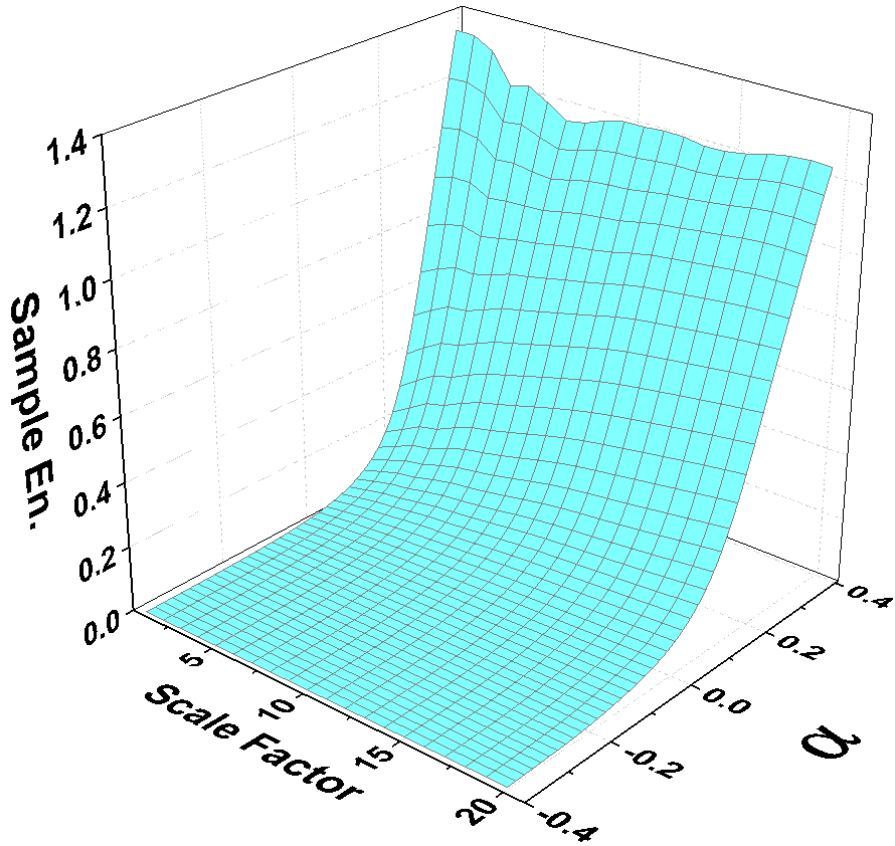

Figure 7: Sample entropy for the cosine series from Equation (14) for $-0.4 \leq \alpha \leq 0.4$, which corresponds to a fractional dimension, $D$, ranging from $1.1 - 1.9$, respectively.

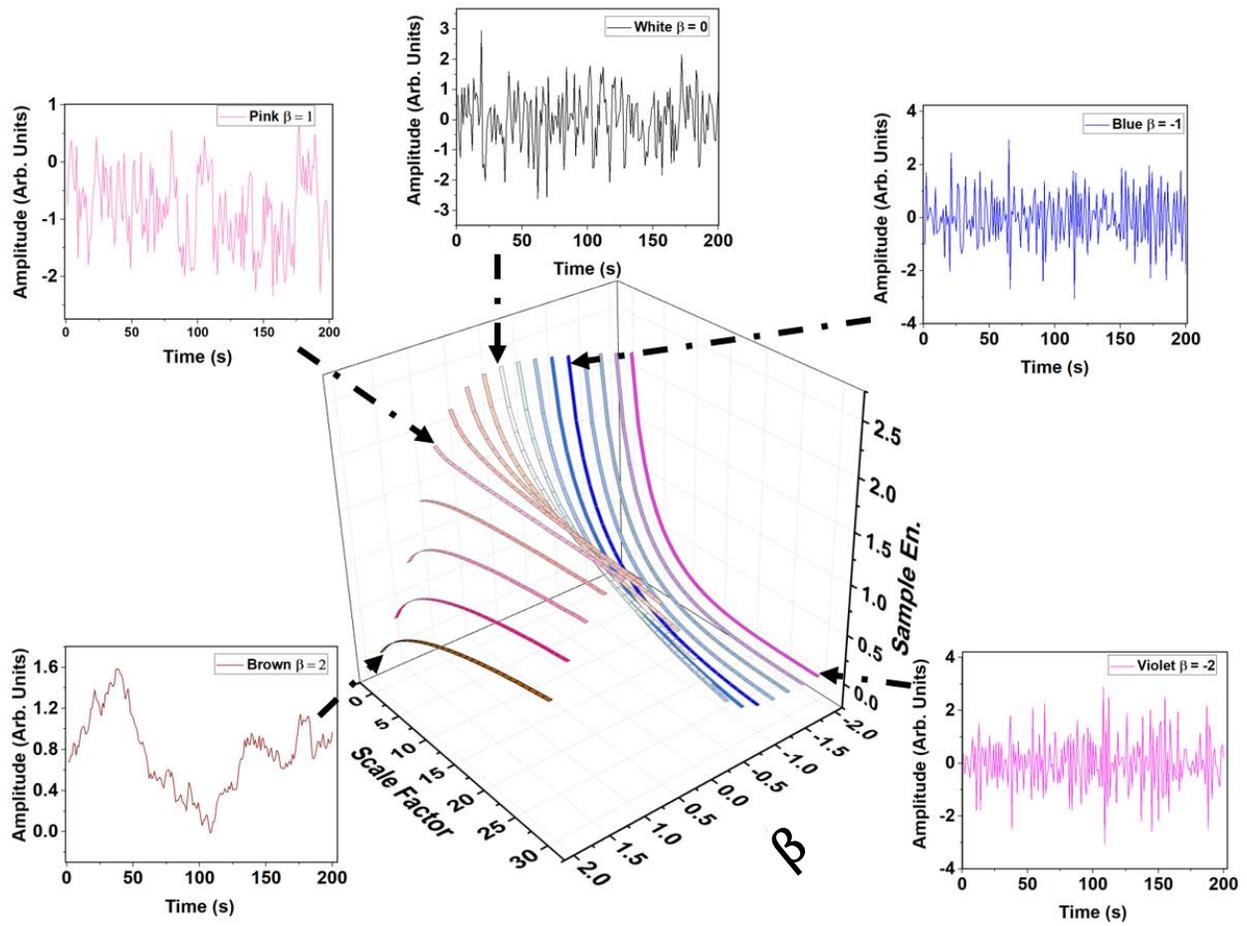

Figure 8: Mean sample entropy of the colored noise data for $-2 \leq \beta \leq 2$ and $1 \leq \tau \leq 30$ plotted with the colored noise plots for brown ($\beta = 2$), pink (or 1/f with $\beta = 1$), white ($\beta = 0$), blue ($\beta = -1$), and violet noise ($\beta = -2$).

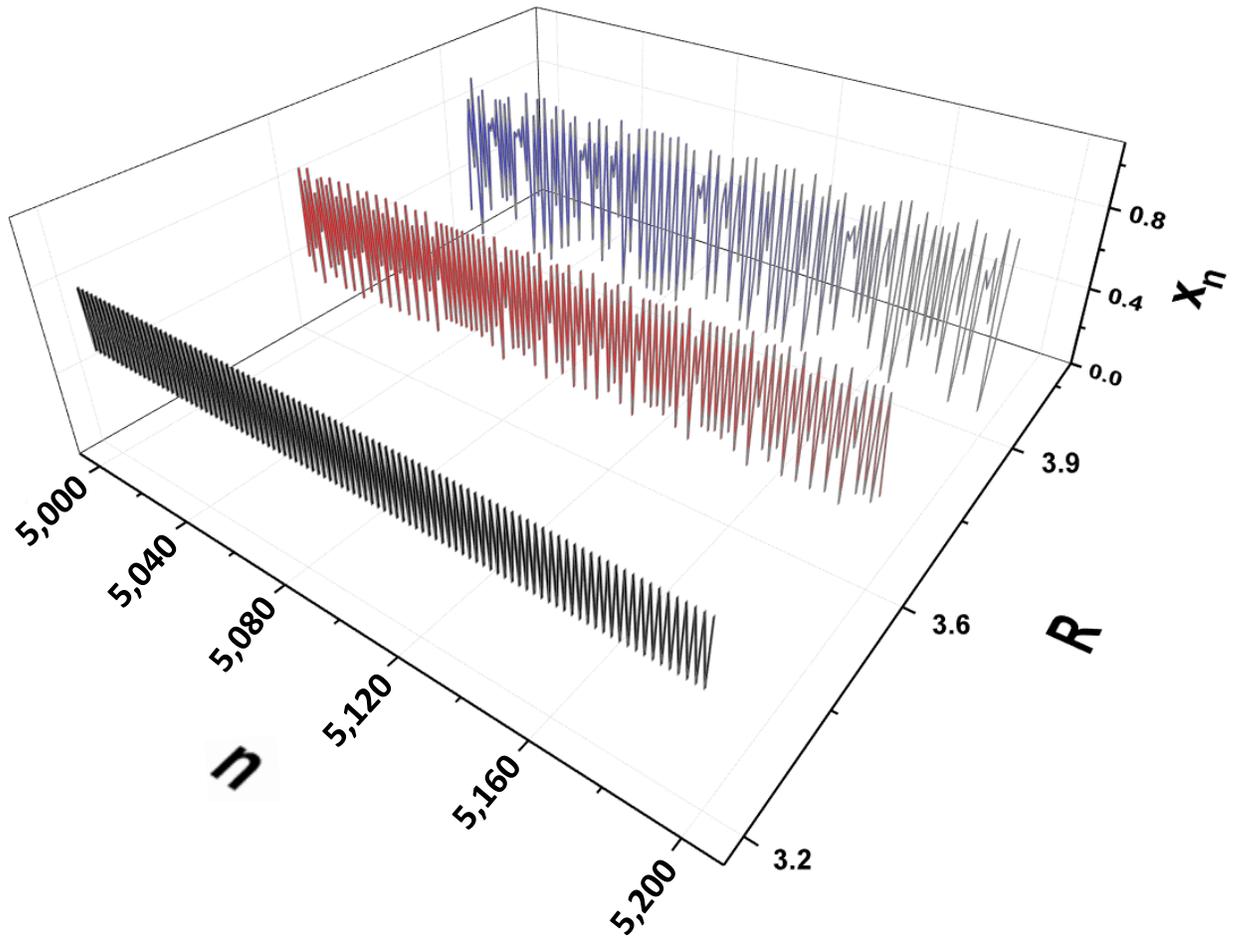

Figure 9. The logistic map with an initial condition of $x_1 = 0.1$, bifurcation parameter, $R = 3.2, 3.6,$ and 3.9 was plotted for $n = 5{,}000 - 5{,}200$.

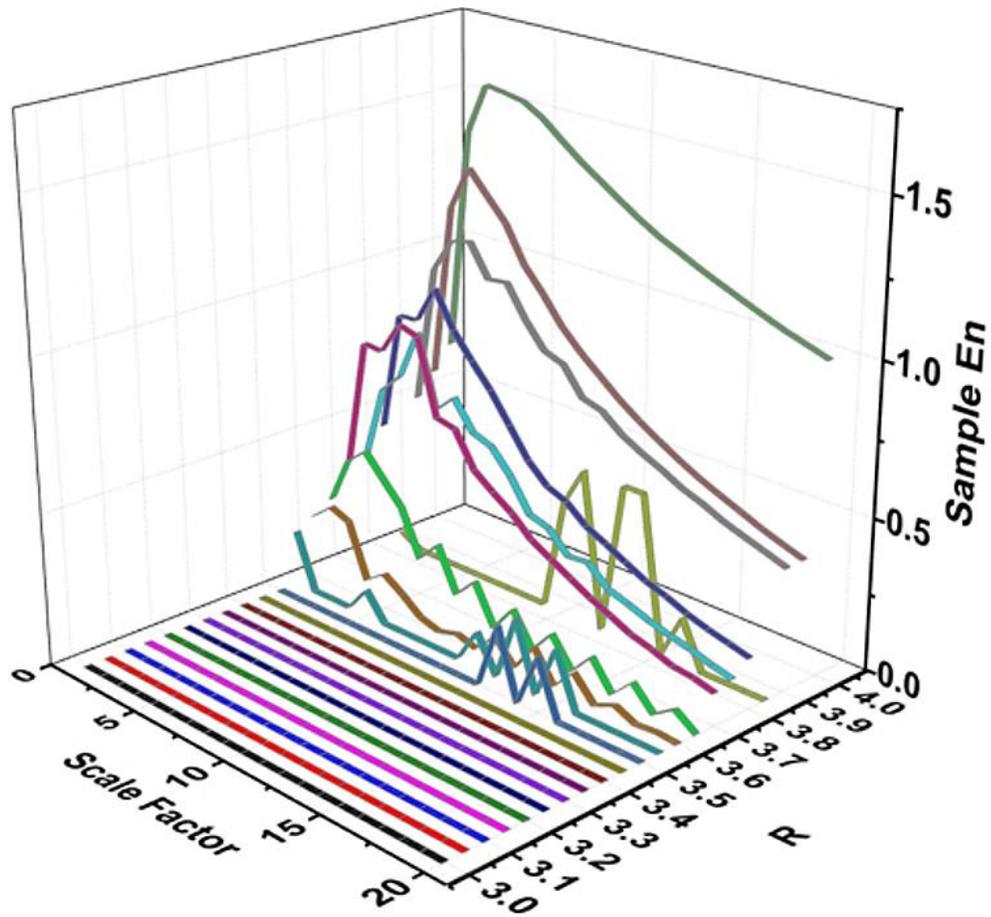

Figure 10. Complexity plot of the logarithmic map for $3 \leq R \leq 4$. Here each sample-entropy curve is plotted for $1 \leq \tau \leq 20$.

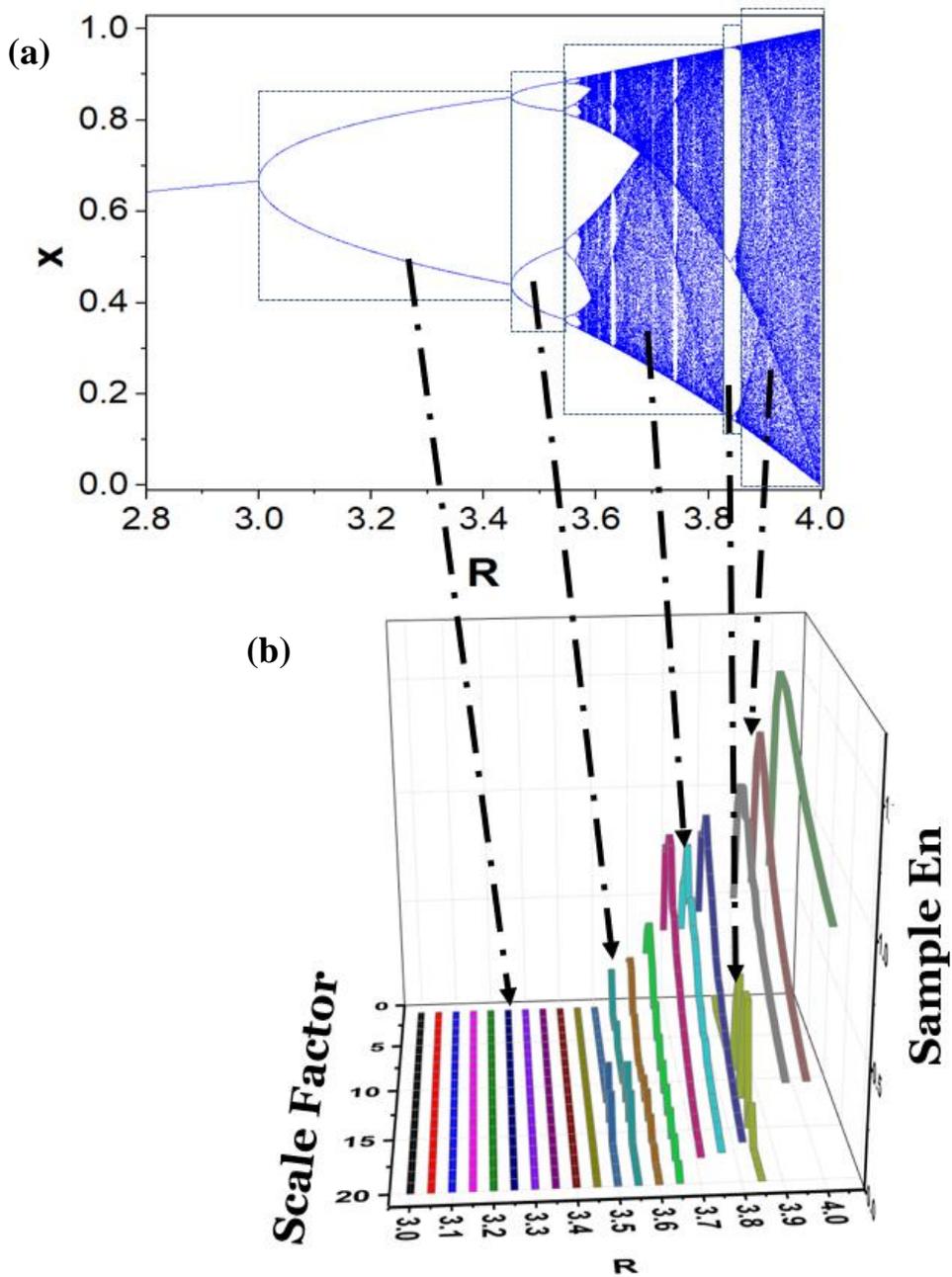

Figure 11. (a)-(b) The bifurcation diagram (plotted for $2.8 \leq R \leq 4.0$) for the logistic map with arrows that link the various regions of the map to its corresponding sample-entropy curves from Figure 10.